\documentclass[onecolumn]{aastex61}
\accepted{November 2016}

\shorttitle{Sports stars: analyzing the performance of astronomers}
\shortauthors{Fluke et al.}

\begin{document}
\title{Sports stars: analyzing the performance of astronomers at visualization-based discovery}
\correspondingauthor{C.J. Fluke}
\email{cfluke@swin.edu.au}

\author{C.J.\ Fluke}
\affil{Centre for Astrophysics \& Supercomputing, Swinburne University of Technology, Hawthorn, Victoria, 3122, Australia}

\author{L.\ Parrington}
\affil{School of Health Sciences, Swinburne University of Technology, Hawthorn, Victoria, 3122, Australia}

\author{S.\ Hegarty}
\affil{Centre for Astrophysics \& Supercomputing, Swinburne University of Technology, Hawthorn, Victoria, 3122, Australia}

\author{C.\ MacMahon}
\affil{School of Health Sciences, Swinburne University of Technology, Hawthorn, Victoria, 3122, Australia}

\author{S.\ Morgan}
\affil{Australian Institute of Sport, Belconnen, Australian Capital Territory, 2617, Australia}
\affil{Computer Science \& IT, La Trobe University, Bundoora, Victoria, 3086, Australia}

\author{A.H.\ Hassan}
\affil{Centre for Astrophysics \& Supercomputing, Swinburne University of Technology, Hawthorn, Victoria, 3122, Australia}

\author{V.A.\ Kilborn}
\affil{Centre for Astrophysics \& Supercomputing, Swinburne University of Technology, Hawthorn, Victoria, 3122, Australia}

\begin{abstract} 
In this data-rich era of astronomy, there is a growing reliance on automated techniques to discover new knowledge.  The role of the astronomer may change from being a discoverer to being a confirmer.     
But what do astronomers actually look at when they distinguish between ``sources'' and ``noise?'' What are the differences between novice and expert astronomers when it comes to visual-based  discovery?   Can we identify elite talent or coach astronomers to maximize their potential for discovery?    By looking to the field of sports performance analysis, we consider an established, domain-wide approach, where the expertise of the viewer (i.e. a member of the coaching team) plays a crucial role in identifying and determining the subtle features of gameplay that provide a winning advantage.  As an initial case study, we investigate whether the {\sc SportsCode} performance analysis software can be used to understand and document how an experienced H{\sc i} astronomer makes discoveries in spectral data cubes.   We find that the process of
timeline-based coding can be applied to spectral cube data by mapping spectral channels to frames within a movie.   {\sc SportsCode} provides a range of easy to use methods for annotation, including feature-based codes and labels, text annotations associated with codes, and image-based drawing.  The outputs, including instance movies that are uniquely associated with coded events, provide the basis for a training program or team-based analysis that could be used in unison with discipline specific analysis software.   In this coordinated approach to visualization and analysis, {\sc SportsCode} can act as a visual notebook, recording the insight and decisions in partnership with established analysis methods.  Alternatively, {\em in situ} annotation and coding of features would be a valuable addition to existing and future visualization and analysis packages.
\end{abstract}
 

\section{Introduction}

Astronomers are now well and truly immersed in a data rich era.  This era commenced around the time that the volume of astronomical data crossed the 100 Terabyte threshold \citep{Brunner02},
and was highlighted by the emergence of the Virtual Observatory [VO; or World-Wide Telescope \citep{Szalay01}] concept.  The VO relied on advances in technology to support access to globally distributed data archives, presenting new opportunities to exploit multi-instrument and multi-wavelength resources [see \citet{Djorgovski05} for the early history].  

As one of the first coordinated, international attempts to federate and serve petabyte-scale scientific data, the VO also played a crucial role in the emergence of a new, fourth paradigm of data-intensive scientific discovery \citep{Bell09,Hey09}.  Data-intensive discovery emphasizes the use of online databases \citep{Szalay06}  and automated data-mining techniques \citep[e.g.][]{Borne09, Ball10, Ivezic14}.
For astronomy, it refers to a state of affairs where the process of discovery moves steadily away from the status quo of the astronomer at a desktop computer working on data stored on a local hard-drive.  Increasingly, it is known as astronomy with ``big data''. 

\subsection{Big Data and Human Discovery Potential}
While a single, simple description of big data in all contexts does not exist, three important characteristics occur in many individual definitions \citep{Ward13}:
\begin{itemize} 
\item Sizes that exceed typical storage capabilities; 
\item Complex structures or relationships within or between the data; and 
\item A reliance on technological or methodological advances for storage, transfer, processing or analysis.
\end{itemize}

Much of the emphasis on big data in observational and survey astronomy centers on the exascale datasets that current [Atacama Large Millimeter Array \citep{Brown04}; Low-Frequency Array \citep{vanHaarlem13}], emerging [Australian Square Kilometre Array Pathfinder \citep{Johnston08}; MeerKAT \citep{Jonas09}], and future facilities [Large Synoptic Survey Telescope \citep[LSST;][]{Tyson02}; Square Kilometer Array \citep[SKA;][]{Dewdney09}] will generate.   These data holdings will need to be calibrated and streamed in real time, they will rely heavily on the use of automated techniques to make discoveries, and they may never be fully analyzed (in comparison with the current level of scrutiny applied to individual observations).    The overwhelming majority of pixels or voxels contained in the survey data will never be looked at by a human.

Due to the dramatic increase in data throughput, there will likely be a corresponding decrease in the time that any one astronomer will devote to looking at individual objects.   An unintended consequence is thsource findingat future astronomers may not benefit from extensive, dedicated time analyzing individual objects.   It is this learning process that allows the current experienced or expert astronomer to have confidence in the interpretation of images (citet{Norris10}; and see Section \ref{sect:expert}).  Simplistically, we need to ensure that we can maximize the human discovery potential when only a limited fraction of the recorded data is visualized, while also providing new training methods for the data-rich/time-poor astronomer.

\subsection{Neutral Hydrogen Spectral Data Cubes}
Consider the specific case of source finding and validation in spectral data cubes, with an emphasis on the search for extragalactic neutral hydrogen (H{\sc i}) at radio frequencies.  We will use this example  throughout this paper, as the status quo comprises automated processes supported by extensive visual inspection and interactive analysis.

A spectral data cube comprises two spatial dimensions (e.g. right ascension and declination) and one spectral dimension. For radio astronomy, this spectral dimension is usually frequency, which is a proxy for the line of sight velocity.     Rest-frame H{\sc i} emits a characteristic signal at 1420.4 MHz, known as the 21-cm line. Factors such as thermal broadening; internal gas dynamics' cosmological and random galaxy motions; and the motion of the Earth shifts the observed frequency from the intrinsic 1420.4 MHz.    Considering that source brightness and angular diameter diminish with distance and recognizing that intrinsic source brightness also varies, it becomes increasingly difficult to distinguish the shifted and smeared extragalactic H{\sc i} signal from the noise in the data cube.  The ideal source finder aims to identify regions of a data cube that are most likely to be a source and not noise or false positives ({\em reliability} approaching $100\%$), while also returning all sources that actually reside in the data ({\em completeness} approaching 100\%).

Fully automated extragalactic H{\sc i} source finding is an unsolved problem \citep{Koribalski12}.  Numerous techniques have been developed, tested, enhanced, and improved [e.g. {\sc duchamp} by \citet{Whiting12};  S{\sc o}F{\sc i}A by \citet{Serra15}; and {\sc SlicerAstro} by \citet{Punzo16}].     Source-finding techniques are often designed or optimised to find the types of sources that are already known about, with the potential that new types of sources are actually rejected. 

There is confidence that the source finders chosen by survey teams will work with a high level of completeness and reliability, but it is not expected that this process will be perfect.   Source finders will miss sources and will likely generate many more false positives than real detections.  Indeed, running different source finders on the same data rarely generates the same set of candidates \citep{Popping12}.    At some point, astronomers will need to look at candidates and make decisions as to whether a ``source'' is real or not.

\begin{figure*}[t]
\begin{center} 
\includegraphics[scale=0.30, angle=0]{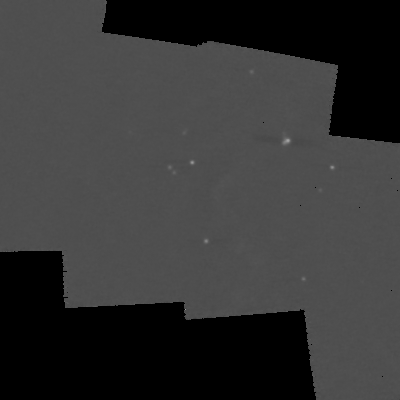}
\includegraphics[scale=0.30, angle=0]{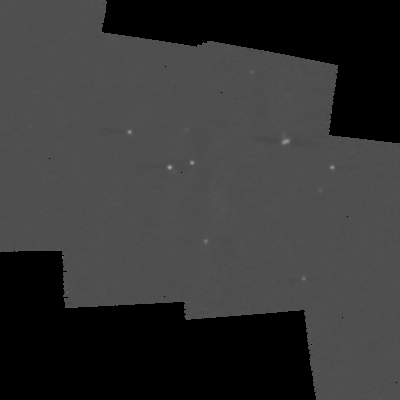}
\includegraphics[scale=0.30, angle=0]{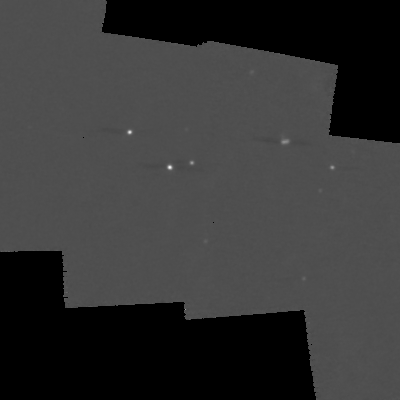}
\includegraphics[scale=0.30, angle=0]{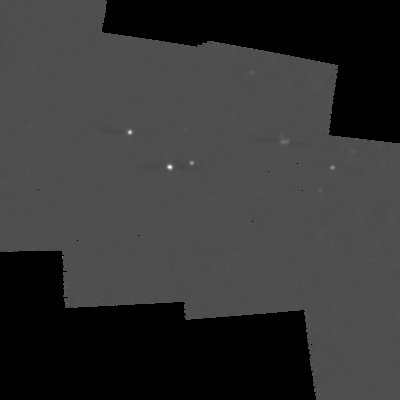}
\caption{Image slicing: four successive channel maps from the Ursa Major region spectral data cube (see Section \ref{sct:ursadata}).  By converting this sequence of channel map images into a movie format, the spectral data cube is ready to be inspected and annotated in {\sc SportsCode}. In this sequence, the sources (white) change in brightness compared to the background (dark gray) noise level. }
\label{fig:slicing}
\end{center}
\end{figure*}

Over several decades, a suite of visualization techniques have been used in extragalactic H{\sc i} studies, including
\begin{itemize}
\item Channel maps, where an individual spectral channel is extracted from the spectral cube and viewed as a two-dimensional image;
\item Moment maps, where an image is created via the projection of data values along the spectral axis, weighted by voxel intensity, velocity or velocity dispersion;
\item Position-velocity diagrams, where a projection occurs along one of the spatial axes; and
\item Volume rendering, which captures the full three-dimensional content of the data cube to show the ``{\em wispy tendrils}'' \citep{Gooch97} of gas.   
\end{itemize}

An extension to displaying individual images along a single axis is the use of imaging slicing \citep[see, for example, ][] {Borkin05}.   Here, slices are extracted from the spectral data cube along orthogonal axes, where they can be displayed side-by-side.  Alternatively, slicing can occur along one axis, generating a frame-sequential set of images that can be converted into a movie (see Figure \ref{fig:slicing}).  

\subsection{Performance Analysis}

What we would like to be able to do is investigate and record {\em how} extragalactic H{\sc i} astronomers currently inspect spectral data cubes and make decisions that a group of pixels or voxels is indeed part of a source and is not an imaging artifact or noise.  The hope is that we may then gain insight into the {\em process} of visualization-based knowledge discovery in this field.   Unfortunately, astronomical software (e.g. {\sc kvis}\footnote{\url{http://www.atnf.csiro.au/computing/software/karma/}},  {\sc visions}\footnote{Part of the {\sc gipsy} package: \url{https://www.astro.rug.nl/~gipsy}}, and {\sc SAOImage DS9}\footnote{\url{http://ds9.si.edu/site/Home.html}}) is rarely designed to deal with documenting this process of selection, classification, decision making, and annotation.    Workflow solutions, such as the VO-oriented AstroTaverna \citep{Ruiz14} or the {\sc encube} visual analytics framework for CAVE2 and desktop \citep{Vohl16}, record the sequence of steps that occurred but not necessarily why specific decisions were made.   Modifying existing software to achieve this aim seems unjustified, so we need an alternative approach.  

In other fields, the situation is very different.  Work practices are established that support the identification, documentation, annotation, and replay of the discovery and decision-making process.   By looking to the field of sports performance analysis, we can learn about an established, domain-wide approach, where the expertise of the viewer (i.e. a member of the coaching team) plays a crucial role in identifying and determining the subtle features of gameplay that provide a winning advantage.   We also start to bring concepts from coaching and talent identification into astronomy, as we begin to explore whether we can identify or train astronomers with elite-level visual discovery skills.

\subsection{This Work}
In this paper, we report on a preliminary trial using sports performance analysis software -- {\sc SportsCode} -- to document and understand the discovery processes occurring in extragalactic H{\sc i} analysis.   {\sc SportsCode} is widely used in coaching; recruitment and talent identification; and injury prevention and rehabilitation processes.    To our knowledge, this is its first use in astronomy.

In Section \ref{sect:vis}, we examine the changing role of astronomers in a real-time, streaming data era and investigate the concept of expertise.  In Section \ref{sect:sports}, we introduce the {\sc SportsCode} software, and report on a preliminary investigation into the coding and annotation of a typical visual discovery process using neutral hydrogen spectral data cubes.   In Section \ref{sect:discussion}, we discuss the strengths and weaknesses of the {\sc SportsCode} experience. We consider how elite talent identification might be required in astronomy.  We encourage nurturing and coaching of all astronomers in order to maximize their potential for visual-based discovery in an increasingly data-rich/time-poor era.

\section{What are We Looking At?}
\label{sect:vis}
Visualization is a crucial component of knowledge discovery in astronomy \citep{Hassan11}. With its foundation in the visible,  astronomy has advanced through its reliance on the human visual system to identify features of interest.  These {\em sources} exist in {\em images}\footnote{For brevity, we use the term {\em image} to encompass conventional two-dimensional pixel-based images and higher-dimensional volumes or hyper-spectral cubes, from a variety of wavelength regimes. } that may have a low signal-to-noise ratio, a high dynamic range, or suffer from significant imaging artifacts (e.g. cosmic rays, over-saturation of CCD pixels, radio-frequency interference and a variety of calibration errors).     At present, humans have pattern recognition and feature identification skills that exceed those of any existing automated approach.

\citet{Rogowitz12} investigated the boundary between human and machine-based discovery, proposing a framework that codifies the variety of visual tasks that are required.  They identify the crucial role played by humans with regards to making judgements about data, whether based on what is seen (perception), what is understood (cognitive), or the aesthetics of the visualization.  The framework is one of continuous human-computer interaction, that starts with the choice about what data to visualize, continues through the selection of relevant visualization algorithms, and allows for interaction with visualizations to modify parameters and enhance discovery.

\subsection{Do we Really Need to Look at Everything?}
As the quantity and quality of image data has increased (e.g. improvements in the number of spatial pixels, spectral resolution, field-of-view and sky coverage), there has been a growing reliance on automated techniques to find and characterize sources \citep{Borne09,Ball10,Ivezic14}.     Taken to its unlikely extreme, the role of the astronomer would change from being a {\em discoverer} -- looking at every pixel on every frame -- to one of {\em confirmer} -- looking only at those pixels identified as interesting by a source finding or catalog-based data mining algorithm.  

More likely, a subset of high probability candidates, including false positives, would be inspected and subjected to further analysis.  The results of this investigation would form part of a feedback loop that works to retrain or improve the automated pipeline, while ensuring sufficient flexibility that entirely new classes of sources can be found.    

The Galaxy Zoo citizen science project \citep{Lintott08} provides an interesting example of the astronomer's changing role. This project addresses a visualization-intensive task - morphological classification of galaxies - for which data sets have grown too large for analysis by experts. Instead, Galaxy Zoo enlisted the human visual system on a large scale, recruiting citizen volunteers to perform galaxy classifications using an online tool. Provided with a minimal level of focused training, these volunteers have provided millions of reliable visual classifications, as well as original discoveries of new classes of object \citep{Lintott09}. However, even this highly successful example of the use of the human visual system is looking toward automation, with machine learning techniques able to reproduce better than 90\% of visual classifications \citep{Banerji10}.

\begin{figure*}[t]
\begin{center} 
\includegraphics[scale=0.4, angle=0]{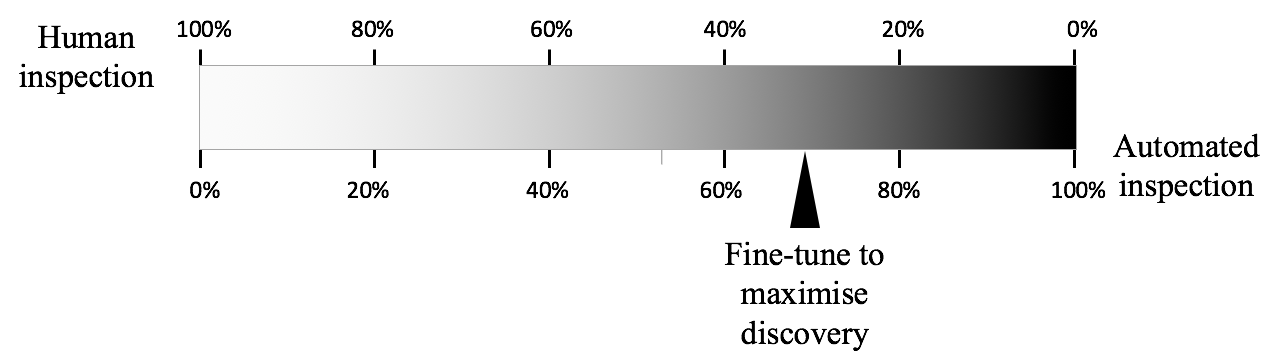}
\caption{There is a fine-tuning necessary between human and automated inspection.  Any particular discovery process must choose where the optimum location occurs. }
\label{fig:finetune}
\end{center}
\end{figure*}

Figure~\ref{fig:finetune} provides a schematic representation of the fine-tuning that occurs between human and automated inspection.   Any particular discovery process must choose where the optimal location occurs in this inspection continuum.   
To maximize throughput it may be necessary to favor the fully automated end of the spectrum.   By viewing fewer pixels, it may not be possible to know what was missed or why certain types of objects are easier or harder for the automated system to identify.
To maximize potential for discovery -- including through serendipity \citep{Fabian09} -- or to emphasize quality control and validation, an increased component of human inspection is likely.   The more time spent visually inspecting the data stream, the longer it is before subsequent processing steps can commence, and the greater the risk that astronomers are just looking at noise, errors, or artifacts.

Resistance to a change in role (from discover to confirmer) is likely to occur if astronomers feel that their personal experience and expertise is not being fully utilized.  Ideally, we want to capture the most important elements of the discovery experience and either support or replicate them {\em in silico} in our automated approaches.   But what is it that astronomers actually look for in their data?   How do they tell the difference between signal and noise?  How do they distinguish between real sources and false positive candidates?  How do they know when they have made a discovery?    Is the ability to make discoveries in data an elite trait that some astronomers inherently do better than others, or is it a skill that can be acquired, developed, or coached?

\subsection{Expertise in Visual Processing Domains}
\label{sect:expert}
There are a number of domains in which visual processing is integral to performance. Medical diagnosis in areas such as radiography and dermatology rely on visual inspection and the use of examples. The ability to identify key features and the knowledge of acceptable variations between similar targets (e.g. pink versus red lesion) are developed over time. Testing of diagnostic skill in medicine supports the position that this is an acquired skill, given gradients of performance across levels of expertise [e.g. learners compared to graduates or generalists compared to specialists -- see \citet{Norman06}]. 

Similar to sports coaching, medical diagnosis also acknowledges the critical role of experiential knowledge \citep{Callary12,Hatala99}. Formal knowledge is a foundation for the experience that provides examples and memorable cases of particular conditions and diagnoses \citep{Norman06}. Moreover, the processes used in reading an X-ray or processing the particular discrete features of an electrocardiograph (ECG) for diagnosis can be influenced by irrelevant accompanying information from prior examples. \citet{Hatala99} showed that ambiguous ECGs with clinical information irrelevant to diagnosis (age, occupation, and gender of the patient) but similar to that of an example item with a different diagnosis were less accurately assessed by residents. Medical students were not influenced to the same extent by the accompanying information, however. These findings support the idea that these types of visual diagnoses use instance-based categorization, in which individual examples are stored, rather than discrete features. In this instance-based processing, new items for diagnosis are matched to the stored examples, in some cases using non-critical information such as age, gender, and occupation. The medical students in this study were less influenced by prior examples, giving insight into the mental structures developed by the residents. Critical to this work is the fact that the processes used for categorization when using instance matching are not accessible through methods such as verbalization in the same way that matching features may be.   

A study similar in design to that used by \citet{Hatala99} showed that visual processing in sport can also be influenced by non-critical information. Baseball umpires were shown borderline baseball pitches (on the border of the strike zone) after being shown pitches that were obvious strikes or obvious balls (clearly outside the target area). The same borderline pitches were rated differently, depending on the context; for example, more ``strike-like'' following an obvious ball pitch \citep{MacMahon08}.  While the influence of contextual information in the case of visual processing in both medicine and sport shows the susceptibility to error, it should be underscored that these errors occur on borderline, ambiguous and difficult items. More importantly, the results show the processes used in categorization, which differ according to expertise level, and develop with experience. These examples also illustrate the usefulness of tracing methods to understand and capture processing, which surely also applies to visual processing in astronomy. 

Processing and decision making in complex problem spaces with multifaceted information sources, the development of skill in these tasks, and the usefulness of understanding how more and less skilled individuals accomplish these tasks is encapsulated in the \citet{Norman06} description of the problem of understanding medical expertise: 
``$\dots$ {\em there are multiple forms of expert knowledge, and each may be used to greater or lesser degree depending on the situation.} $\dots$ {\em Experts (and novices) may invoke causal knowledge, rules relating features to diagnoses, or prior examples to solve the problem. The better question is `How are these various forms of knowledge used in solving clinical problems?'}'' \citep[][page 346]{Norman06}.
This same final question can be applied to understanding human inspection in the discovery process in astronomy to better inform automated inspection.

\begin{figure*}[t]
\begin{center} 
\includegraphics[scale=0.65, angle=0]{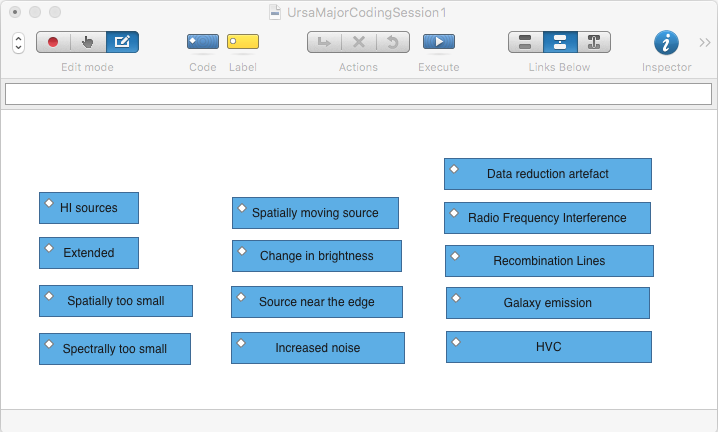}
\caption{The {\sc SportsCode} coding window.  The user defines a set of codes and labels that will be used to annotate the action presented in the timeline.   These codes were generated before the coding trials were performed. }
\label{fig:coding}
\end{center}
\end{figure*}

\begin{figure*}[t]
\begin{center} 
\includegraphics[scale=0.45, angle=0]{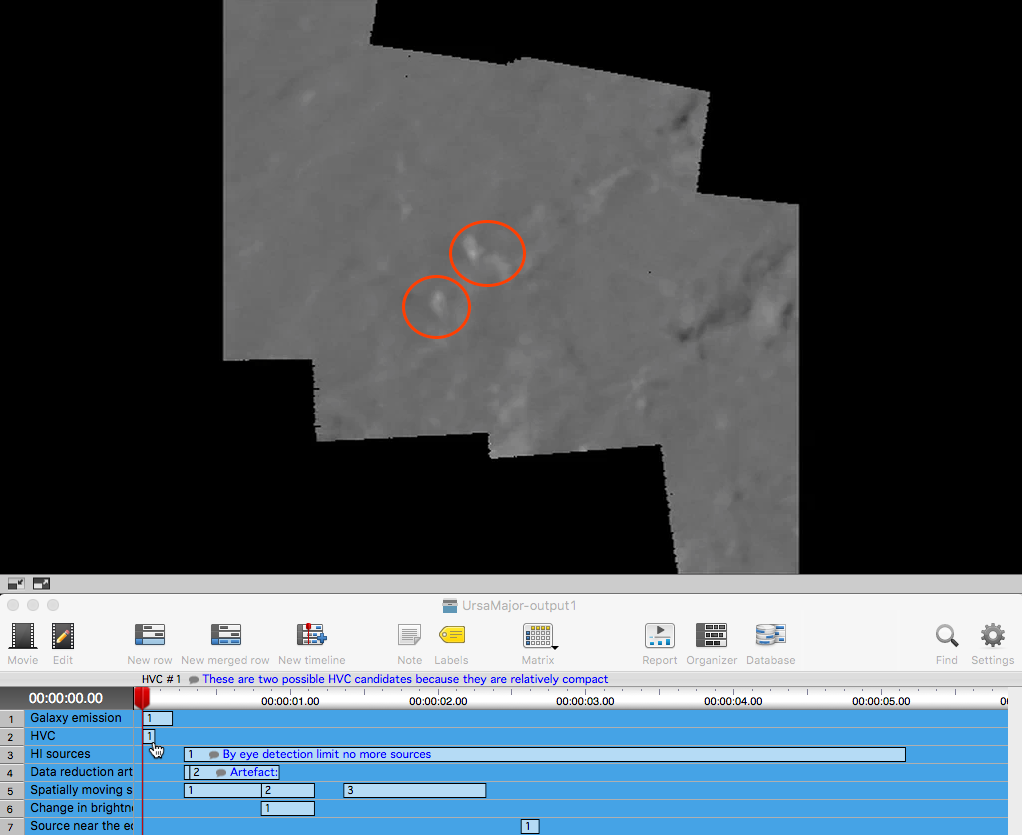}
\caption{ A {\sc SportsCode} timeline. Top: the movie currently under review.  Bottom: the timeline window where codes and labels (not used in this work) are applied to the movie.  Each code occurs in a different row in the timeline, with individual instances appearing as extended colored rectangles.  Each instance of a code is assigned a unique number.  In this example, two red circles have been drawn in the movie window indicating the possible presence of high-velocity clouds, and a note has been added to the coded region (HVC \#1). }
\label{fig:timeline}
\end{center}
\end{figure*}

\section{{\sc SportsCode}: Performance Analysis in Action}
\label{sect:sports}
The goal of any performance analysis process is to understand, identify, and improve those facets of an activity that lead to a winning advantage.  Coding image sequences (i.e. movies) is a common process for knowledge discovery in sport performance analysis.   Looking for specific instances of, for example, successful and unsuccessful gameplay allows coaching staff and players to obtain insight into how individual and team-based actions contribute to the outcome.   In a real-time mode, coding may lead to a modification of current tactics in order to regain or maintain the lead.   In a post-game mode, individuals and teams can be debriefed on areas of achievement and aspects for improvement and further coaching or skills development.

{\sc SportsCode} is part of a suite of commercial annotation packages developed originally by Sportstec, and included in the suite of sports performance tools offered by Hudl\footnote{\url{http://www.hudl.com}}.   It has been used to support knowledge discovery in sports ranging from hockey and basketball to Australian Rules Football and cricket.   {\sc SportsCode} can be used in a real-time mode, taking input from multiple live-streaming (i.e. broadcast) video sources, or for post-game analysis, by loading saved movie files.   

{\sc SportsCode} is generally used for two distinct tasks.  The first task is the labeling of discrete events (such as a goal being scored) for subsequent non-linear analysis by a coach.  We use non-linear in a time-based sense: the coach will not re-watch the entire game from end to end, but will focus on the specific actions or events of importance.   The second task is labeling of more subjective game events that have some meaning to the coach.  This might include particular set plays that the coach has established with the team, or evidence of a pattern of behaviors that starts to emerge as players get fatigued.    In both labeling scenarios, the value of {\sc SportsCode} is that it allows the coach to compile sets of possibly sparse events, and then navigate rapidly to events of interest.    Additionally, through the use of consistent labeling across multiple games, tournaments or seasons, a comprehensive library of actions is gathered.   This can be used to provide ``highlight reels'' for individual athletes or to help plan improved strategies for the team.  An area of interest is to use the labeling as a training set for an intelligent system to be able to recognize the sequences as they occur, however that is beyond the scope of the present work.

\begin{figure*}[t]
\begin{center} 
\includegraphics[scale=0.5, angle=0]{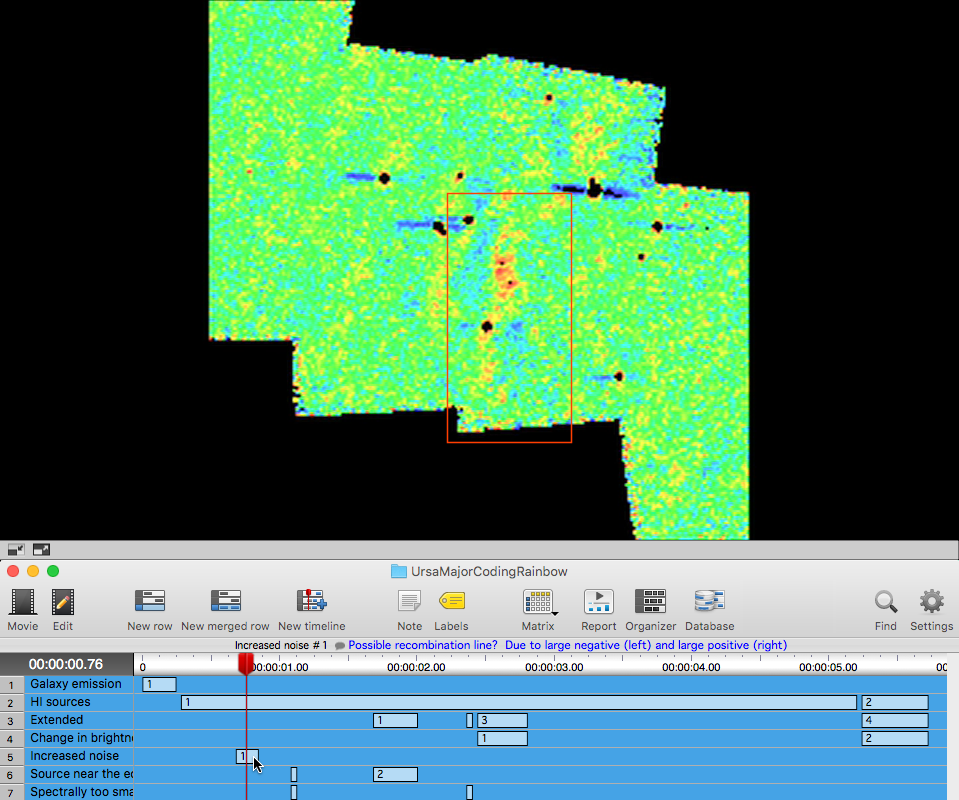}
\caption{During coding of the UM2 movie, a feature was identified in channels 19-21 (one frame either side of timecode 00:00:00.76).  This feature was coded as {\em Increased noise}, with a note that it was a ``Possible recombination line? Due to [the appearance of a] {\em large negative (left) and large positive (right)} [change in flux]'', as highlighted by the red rectangle. }
\label{fig:recombination}
\end{center}
\end{figure*}

\begin{figure*}[t]
\begin{center} 
\includegraphics[scale=0.5, angle=0]{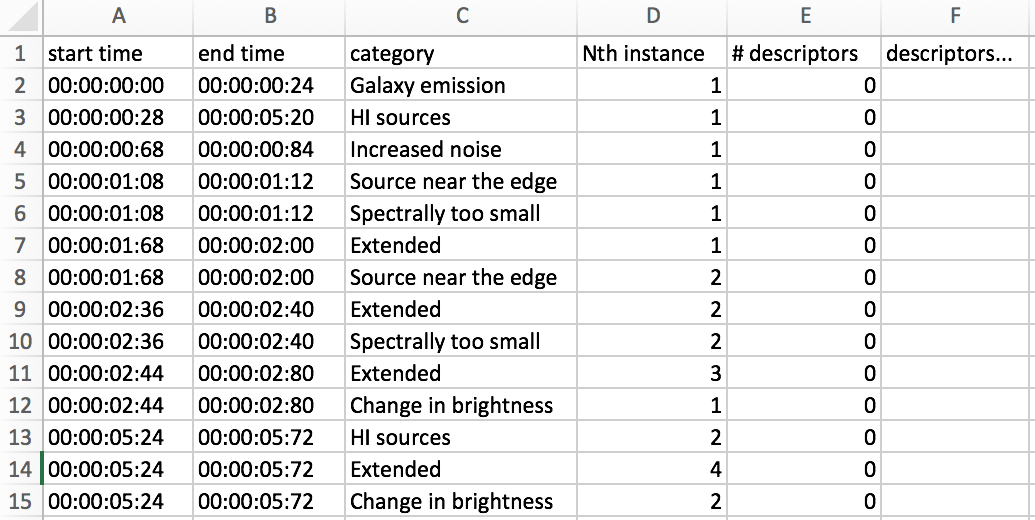} 
\includegraphics[scale=0.5, angle=0]{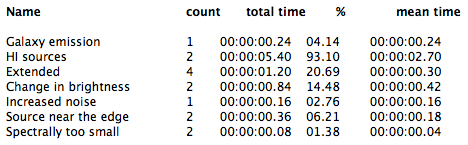}

\vspace{0.3cm}
\includegraphics[scale=0.3, angle=0]{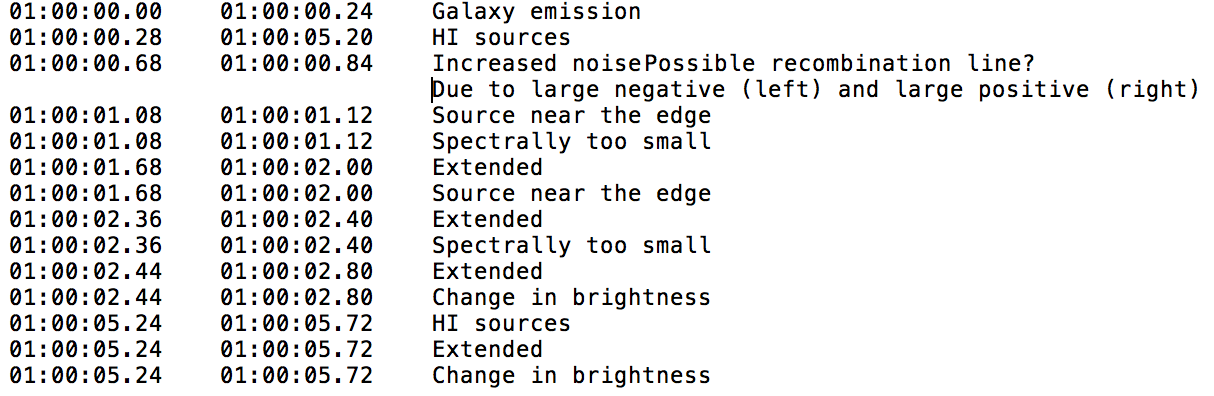}

\caption{Three non-graphical outputs available after coding include (top left) an Excel edit list;  (top right) an instance frequency report, which summarizes the proportion of the timeline containing each code; and (bottom) a timecode log containing additional notes recorded during the coding process.}
\label{fig:outputs}
\end{center}
\end{figure*}

{\sc SportsCode} provides customisable feature-based selection and annotation of events or event sequences ({\em Codes}; see Figure \ref{fig:coding}), which are integrated within the {\em Timeline} for playback, export or further analysis (Figure \ref{fig:timeline}, bottom).  Functionality is provided for {\em Drawing} on frames, including circles, rectangles, free-hand lines and text annotation (Figure \ref{fig:timeline}, top and Figure \ref{fig:recombination}).     The outputs from {\sc SportsCode} are aimed at performance analysts.  For text-based data this occurs in Microsoft Excel or comma-separated variable (CSV) formats (Figure \ref{fig:outputs}), while image-based data can be exported as short movie sequences (associated with a specific code, and including any graphical annotation) or individual frames.

{\sc SportsCode} is currently only available for Apple Mac computers; in this work, we used {\sc SportsCode} Version 10.3 on an Apple 
iMac.\footnote{21.5-inch iMac, 2.9 GHz Intel Core i5 with NVIDIA GeForce GT650M, El Capitan (Version 10.11.5)}   Alternatives to {\sc SportsCode} include packages such as Dartfish\footnote{\url{http://www.dartfish.com}} and Nacsport\footnote{\url{http://www.nacsport.com}}.   These options were not investigated: our purpose is to consider whether sports performance analysis tools can be used to understand how astronomers make discoveries in data -- not whether {\sc SportsCode} is the best environment to achieve this goal.

In the remainder of this section, we report on our experiences using {\sc SportsCode} as a framework for understanding how extragalactic H{\sc i} astronomers make or validate discoveries.  We are not suggesting that {\sc SportsCode} is an appropriate solution for performing qualitative or quantitative visualization and analysis for spectral data cubes.  Many other solutions exist which are specifically targeted at astronomy data in general and spectral data cube visualization in particular.   See \citet{Hassan11} for an overview of spectral cube visualization, and the more recent work by \citet{Hassan13},  \citet{Kent13}, \citet{Perkins14}, \citet{Punzo15}, \citet{Taylor15}, \citet{Naiman16}, and \citet{Ferrand16}.    However, certain functionality---particularly the ability to both annotate and draw regions onscreen that were linked to events in the timeline---was found to provide a simple, yet potentially powerful, addition to existing solutions.

\begin{figure*}[t]
\begin{center} 
\includegraphics[scale=0.5, angle=0]{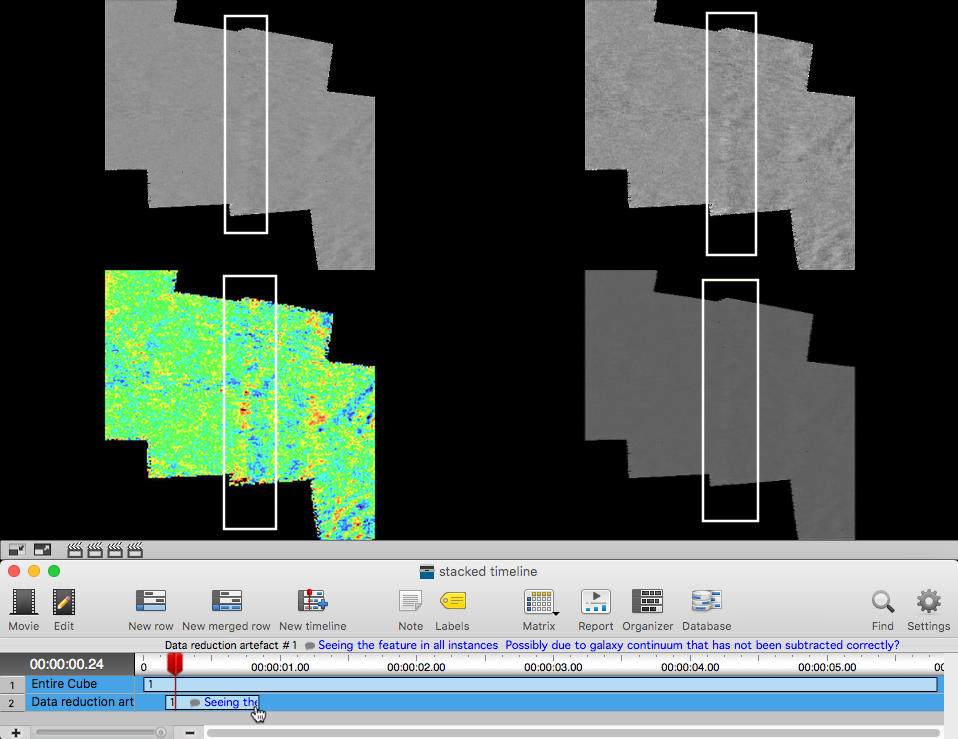}
\caption{A {\sc SportsCode} stacked instance movie.  In this view, {\sc SportsCode} supports multiple movies that are played back synchronously.  The featured highlighted by the white rectangles was originally coded as a {\em Data reduction artifact} during inspection of UM1.  The possible cause was an error when the original sub-cubes had been merged to create the Ursa Major region data cube.   When viewed in this stacked mode, the diagnosis was changed as being more likely an issue in the way the galaxy continuum was originally subtracted.  }
\label{fig:allscreens}
\end{center}
\end{figure*}

\subsection{Preparation}
\label{sct:ursadata}
For our trial we use a spectral data cube from the Ursa Major region, a field with an atypically rich population of spiral galaxies.  This data cube was obtained during the H{\sc i} Jodrell All-Sky Survey \citep[HIJASS;][]{Lang03}, a blind H{\sc i} survey conducted with the 76.2-m Lovell telescope at the Jodrell Bank Observatory.  The Ursa Major region spans right ascensions from 10$^h$45$^m$ to 13$^h$04$^m$, and declinations from 22$^\circ$ to 54$^{\circ}$.  The data cube comprises the combination of seven individual HIJASS data cubes.   Full details of the Ursa Major region study may be found in \citep{Wolfinger13}.

{\sc SportsCode} is not image processing software, and does not permit {\em in situ} manipulation of the data range, color maps, etc.   Moreover, it is designed to work with broadcast and online movie formats (such as the Apple MOV and the closely related MP4\footnote{An implementation of the ISO/E 14496-12:2004 standard} media container formats).  Consequently, a format conversion step from an astronomical data format (e.g. FITS) to MOV or MP4 is required.   For the purpose of this investigation, this conversion represents a minor overhead.   

Using a pre-existing, custom S2PLOT \citep{Barnes06} code, we assign a color map of red, green, blue (RGB) values to the  voxel data values and export each spectral channel as a separate image.  For reasons of convenience, this initial export is performed in PPM format\footnote{``A lowest common denominator format'': \url{http://netpbm.sourceforge.net/doc/ppm.html}}.  The ImageMagick\footnote{\url{http://www.imagemagick.org}} {\sc convert} application is used to convert frames from PPM to the TARGA/TGA\footnote{See for example \url{http://www.fileformat.info/format/tga/egff.htm}} format.   Next, the image sequence is  read into Apple Final Cut Pro\footnote{\url{http://www.apple.com/final-cut-pro/}} and exported as a MOV file.   We recognize that this is an overcomplicated process!  The point here is that we are producing a movie that can be opened in {\sc SportsCode} using tools that were available on-hand, rather than attempting to develop a generic {\sc fits\_to\_mov} application.   There is likely to be some degradation in this multi-format conversion process, but we are not using {\sc SportsCode} to analyze the resultant data.   Overall, this approach gave us more control over the data range and color map than we could achieve by reading a FITS cube into SAOImage DS9 and exporting directly as an MPEG\footnote{\url{http://mpeg.chiariglione.org}} movie file.  In future work, we hope to simplify, and more easily automate, this conversion process.

The Excel and CSV-formatted {\sc SportsCode} outputs (Figure \ref{fig:outputs}) provide timestamps (hh:mm:ss.ss) from the timeline for the start and end of codes and labels.  Matching a spectral channel to a time-stamped frame requires a sensible choice of the frame rate when the movie is encoded.    For simplicity, we chose 25 frames per second (fps), or 0.04 seconds per frame: a spectral data cube with 100 channels will have a duration of 4 seconds.    While this leads to rapid playback in the timeline, frame-by-frame stepping is supported via the arrow keys.   It is straightforward to convert a time-stamp back to a channel number for further investigation and analysis.

No attempt was made to account for distortions that would align the Ursa Major data cube with the world coordinate system (RA and Dec).  The cube is rotated 90 degrees counter-clockwise compared to the more conventional alignment of the coordinate axes.  The size of the input data cube was 437 $\times$ 512 spatial pixels $\times$ 145 channels (heliocentric velocities from 300--1900 km s$^{-1}$).   The cube is adjusted to $400\times400\times145$ voxels (extracted as a sub-cube with an image aspect ratio of 1:1 for convenience) and flux values are rescaled from $0.0$--$1.0$ in arbitrary units (see Table \ref{tbl:values} for the resultant data ranges).  The output from Final Cut Pro is contained within a movie frame with dimensions $1440\times1080$ pixels, using the HDV-1080p25 sequence pre-set.      None of these decisions has an impact on our use of the cube within {\sc SportsCode} -- we are not attempting quantitative work.   

Four movies were generated and inspected: 
\begin{itemize}
\item UM1 uses a linear mapping of voxel values to grayscale pixels, covering the full range of the normalized flux values;
\item UM2 uses a linear mapping of voxel values to a rainbow color map (from blue to red) over a reduced range. This sequence emphasizes voxel values near the mean value from the cube.   This choice exaggerates some features, while likely hiding others.  High-flux sources are truncated, and appear black in the frames;
\item UM3 uses the same data range mapping as UM2, but this time with a grayscale color map; and 
\item UM4 uses a data range intermediate to UM1 and UM2, with a grayscale color map.
\end{itemize}

\begin{table*}[h]
\begin{center}
\caption{Ursa Major Region sequences used in the three stages of coding trials.}
\label{tbl:values}
\begin{tabular}{lrrcll}
{\bf Sequence} & {\bf Min} & {\bf Max} & {\bf color map} & {\bf Trial} & {\bf Notes}  \\
\hline
UM1 & 0.0 & 1.0 & grayscale & 1 & This data cube is dominated by sources  \\
UM2 & 0.24 & 0.38 &  rainbow (blue to red)  & 2 & This data cube shows more noise and artifacts\\
UM3 & 0.24 & 0.38 &  gray & 3 & Used as part of the stacked timeline movie trial \\
UM4 & 0.14 & 0.48 &  gray & 3 & Used as part of the stacked timeline movie trial \\
\hline
\end{tabular}
\end{center}
\end{table*}

\subsection{Codes and Coding}
Coding was performed on 2016 October 26 in a single 90-minute session.  Present at the session were Christopher Fluke (C.F.; overseeing the use of {\sc SportsCode}), Virginia Kilborn (V.K.; the expert H{\sc i} astronomer), and Sarah Hegarty (note taking of the discussion occurring between C.F. and V.K. during the session).   C.F. started the coding session with a general introduction to {\sc SportsCode} and the goals of the coding session: to establish whether the software could be used to understand how V.K. inspected and made decisions about the nature of objects within the Ursa Major region spectral cube.  

Through discussion with, and questioning by C.F., V.K. suggested a set of codes that represented the typical features expected in an H{\sc i} spectral data cube.  These included features relating to the spatial and spectral  attributes of sources, channel-based changes in brightness and spatial position, and non-source characteristics, such as lack of spatial or spectral extent, signal associated with the Galaxy or high-velocity clouds (HVCs), or data reduction artifacts.   In general, the features that indicate the presence of a sources are a spectrally extended structure (i.e. present in more than one frame), with a spatial extent that was at least as large as the observing beam (a few pixels across).   Internal motions, particularly coherent rotation of an H{\sc i} disk, cause the brightness and spatial position to vary from frame to frame.   Structures that were not real, such as Galaxy foregrounds, are sometimes hard to disentangle from real emission at low velocities (the early frames in the movies), as line of sight velocity does not necessarily indicate distance, particularly for HVCs. These were entered as codes via the coding window (see Figure \ref{fig:coding}).

A new timeline was created, and the UM1 movie was loaded (Trial 1).   C.F. demonstrated how a code is applied by activating and then deactivating one of the coding buttons, which results in the creation of a coded Instance in the timeline window (see Figure \ref{fig:timeline}).    V.K. was asked to look through the UM1 movie and nominate when particular features were seen that matched the pre-established codes.   An exhaustive attempt to find all instances of each code was not attempted.  Instead, we considered whether there was evidence that a subset of the codes could be matched to features.  As the UM1 covers the entire voxel-value range, bright sources stood out clearly against a relatively smooth background.  Spectral features relating to the Galaxy were obvious in the first five frames.

Multiple passes through the dataset were required to code for different features.    In many sport scenarios, there are mutually exclusive periods of play or clear isolated classifications (e.g. in tennis, the ball is either in the server's side of the court or the receiver's side of the court; it cannot be in both at the same time).   {\sc SportsCode} allows these sequences to be linked together so that activating a code button relating to one phase of play deactivates one or more alternative codes.   This is not necessarily the case in extragalactic H{\sc i} astronomy.  If each potential source was coded with start and stop times, there would be a great deal of overlap as multiple, distinct sources are visible in each frame.

A feature that immediately caught V.K.'s attention was the ability to annotate and draw on frames.  Even a simple annotation, such as a circle or rectangular bounding box, drawn onto a frame helps to focus the user's attention.  Additionally, these image-based annotations can be animated, so that they appear and disappear at user-defined key frames within the movie.   Drawings and other onscreen annotations can be saved, and even exported in an embedded form as an Instance Movie.

The same process was then performed for UM2 (Trial 2; see Figure \ref{fig:recombination}).   More effort was expended on looking for features that had not been coded during the inspection of UM1.  Again, the coding attempt was not exhaustive.  With a restricted portion of the data, examples of a radio recombination line and a source near the edge of the observing field were more easily visible.

The final coding trial (Trial 3) utilized the stacked instance movie feature.  A new timeline was created for each of the four movies. A single ({\em Entire Cube}) code was applied to the individual timelines, which allows the sequences to be converted into Instance Movies.   The four new Instance Movies were then combined into a stacked timeline movie (Figure \ref{fig:allscreens}).  The movies can now be coded together, as the playback of frames is synchronized.  It is possible to toggle display of each of the  individual instances.

The need to pre-process and encode spectral cubes as movies means that there is no option for interactive control during the coding process. Instead, the ability to use a stacked instance movie was instructive, as now four movies with different data ranges and/or color maps were available in a single coding process.   This allows the expert astronomer to make comparisons as part of the process of identifying or validating sources and non-sources.  

During inspection of UM1, V.K. coded a feature in channels  as being due to a data reduction artifact: a large vertical feature was thought to be an issue when sub-cubes were joined to created the complete Ursa Major cube.   Reviewing the stacked instance movie, V.K. changed her mind about this feature, deciding it is more likely to be continuum emission that had not been subtracted correctly. Visualization can be very influential over findings, even for an expert!

\section{Discussion}
\label{sect:discussion}
The increased reliance on data mining and other automated techniques changes the way that astronomers undertake one of their key roles: spotting interesting things (discoveries, anomalies, artifacts, relationships, and trends).   In this work, we have investigated whether performance analysis software developed for coaching and development of elite sportspeople can help us to understand the way that astronomers make discoveries in data.    To provide the most useful visualizations of the most relevant data, we need to understand what astronomers are actually looking for.

\subsection{The {\sc SportsCode} Experience}
Once the necessary conversion of a FITS-format spectral data cube to a movie format was made, {\sc SportsCode} was straightforward to use.    We reiterate that {\sc SportsCode} is not intended to be used for qualitative or quantitative data analysis tasks.   Its purpose is to allow timeline-based annotations and frame-based drawings to be supported and exported for further analysis.     For our example spectral data cube of extragalactic H{\sc i} in the Ursa Major region, it was easy for an inexperienced {\sc SportsCode} user -- but expert H{\sc i} astronomer -- to apply a set of pre-defined codes.

It quickly became evident that one of the most important features of {\sc SportsCode} was the ability to draw on an individual frame, associating a visual representation with the Code in the timeline.   Being able to point out a feature, clearly indicate its position, and control when the drawing appeared or disappeared made it easier to go back and review why a decision had been made.  This simple image-based annotation would be a valuable addition to existing and planned visualization and data analysis packages in astronomy.  It provided a valuable aid to memory, while also serving as a potential conversation starter for a training process.

When examining any visualization technique, it can be tempting to ask for more.  Although we did not test it, it would be relevant to look at alternative orientations or color mappings applied to the spectral data cube.  For example, position-velocity movies could be made by slicing along the right ascension or declination axis.   A logarithmic scaling is often a more practical choice, as it reduces the number of color values applied to the strongest sources, so that more of the displayed dynamic range is in the most interesting ``sources just hidden in noise'' regime. These alternatives could be incorporated into a single coding session through the use of a more comprehensive stacked instance movies.   The comparison with sporting analysis is the ability to take input from multiple broadcast cameras, each following the action (or individual players) from a range of different perspectives.  

\subsection{Integration with Astronomical Data Analysis}
A careful choice of color maps for the underlying data (e.g. grayscale) and any {\sc SportsCode} drawings would allow post-processing of exported sequences.   A color-based bounding box can be obtained and then matched back to the original pixel coordinates from the spectral data cube.  Combining this with the frame time stamps, which map to the spectral channels, then a three-dimensional sub-region can be defined and used as an input to an appropriate data analysis package. Alternatively, regions could be used as the input to an image-based automated source finder, such as a deep learning network, as examples of sources, artifacts or noise.  Indeed the reverse source finding process---finding the regions in a spectral data cube that are noise---may prove to be a practical option using deep learning networks, as there are no shortage of source-free, noisy images that can be generated as a training set.

Annotation tools may have an important role to play in supporting validation of existing automated source finders.  While the completeness and reliability of competing source finders can be assessed quantitatively \citep[e.g.][]{Popping12}, embedding source finding outputs within the spectral channel videos would provide an alternative method to judge the success of a source finder.  Of particular note are the ability to understand and document:  sources that are easy to find; cases where multiple components of a single source are identified as separate candidates; issues that may arise near the edges of a spectral data cube; or types of sources that are readily apparent by eye yet the automated technique overlooks.

As the \citet{Rogowitz12} framework demonstrates, the human observer makes on-going judgements about the contents of a dataset.  This can include selecting and applying multiple visualization techniques to the same data in order to establish the necessary insight.   For our case, discovery in H{\sc i} spectral data cubes is not confined to reviewing channel maps.  Extracting a spectrum from the data cube is one the key ways to discriminate between sources and non-sources. 

During the coding session, the possibility of running a discipline specific analysis packages (i.e. {\sc kvis})  alongside {\sc SportsCode} was raised.   A feature that was easier to identify in the timeline-based version could then be matched back to the original data for validation.  This could be done as either a training exercise, to build domain experience, or to evaluate the outcomes of an automated source finding process.   Again, the ability to easily add graphical and text annotations to the channel-based timeline was considered an asset.  In this coordinated approach to visualization and analysis, {\sc SportsCode} can act as a visual notebook, recording the insight and decisions in partnership with established analysis methods.   

This approach does produce some undesirable overheads, in particular the need to convert astronomical data formats into one or more movie files. 
Moreover, as {\sc SportsCode} is commercial software that is currently only available on Apple Mac computers, it is not easily available to all astronomers.   A preferred alternative would be to integrate annotation capabilities directly into existing (where possible) and future visualization and analysis software for astronomy, possibly supported by a flexible and extensible annotation standard.

\subsection{Training and Teamwork}
Science, like sport, is a collaborative, team-based activity.   The success of the team depends on the contributions of each team-member.  The typical astronomical survey team will comprise experienced astronomers, emerging researchers, and rookies (graduate and undergraduate students).  Team members learn from each other, while continually practicing and developing their individual skills, but rarely (if ever) with a formal team coach focused solely on performance.  

Through our investigation of {\sc SportsCode}, we can see how it might be used as part of a training - and coaching - program for new H{\sc i} astronomers.  The aim here is to address the reduced time available to gain experiences at visual-based discovery and validation \citep{Norman06,Norris10}.   The experienced (i.e. expert) astronomer would establish a set of codes that would be used for visual inspection and analysis.  A suitable collection of data would be gathered and prepared for use in {\sc SportsCode}, and this data would be coded as thoroughly as possible by the expert.   Instance-based movies would be generated for a subset of examples of each of the codes and made available to the learner, so that they can start to build an internal mental library of instance-based categories for future `diagnosis' \citep{Hatala99}.   This also provides a reference library that can be used for revision, if required.  Next, the learner would be set a task to code the same data, without access to the codes applied by the expert.  At the end of the coding exercise,  the learner's codes would be compared with the experts codes.   This can be achieved by using the merge timeline capability\footnote{This feature combines all codes from multiple timelines.} in tandem with the Stacked Instance movie option.    It is then 
straightforward to see which codes match, and where any important discrepancies are occurring.  This process may even uncover aspects of the original coding that the expert missed, particularly if the learner has great potential for ``elite'' visual discovery performance.

\subsection{Elite Talent Identification}
The ability to understand or gain insight from a data visualization is a skill.  Through repeated use and directed development (i.e. coaching), it is expected that an individual's capabilities at making a discovery could be improved.   

Treating astronomers as people with different capabilities is not a new idea [see \citet{Berlucchi11} for examples demonstrating the relationship between astronomy and the development of visual neuroscience].   By the early nineteenth century, the {\em personal equation} was in use to account for differences in the reaction times of individual astronomers when performing astrometrical tasks \citep[e.g.][]{Mitchel58,Hollis38}.
The personal equation recognized that some astronomers have faster reaction times than others.   

We suggest that some astronomers may be better than others at using specific visualization techniques (channel maps, image slicing, or volume rendering) or types of displays (two-dimensional and three-dimensional displays, large-scale immersive facilities such as the CAVE2, or the new breed of virtual reality head-mounted displays) to make discoveries.   This may also be dependent on the the processing methods to prepare the visualization, or the nature of the data itself (e.g. a noisy spectral data cube compared to a well-calibrated cube with bright sources). So how do we identify and develop elite talent?
     
A major area of investigation in sports science is talent identification, confirmation, and development. In many ways, it is the `holy grail' of sports science: selecting the appropriate athletes early will theoretically mean the effective use of resources and the conversion of potential into performance and outcomes (e.g. Olympic medals, World Cups). A myriad of scientific and applied sport-specific frameworks have sought to encapsulate this process. The assessment of physiological capacity, cognitive skills and temperament are common, with many sports incorporating selection camps to collect quantitative and qualitative performance data. These data are complemented by in-game performance statistics and season-long scouting reports. 

The use of frameworks such as the expert performance approach \citep{Ericsson93} help to identify the particular key differences between elite performers and their novice cohorts. For example, experts consistently show superior perceptual-cognitive skills, such as the ability to recall patterns of play and choose the best option in a decision-making scenario, but not generic abilities like reaction time to non-domain specific stimuli. The development of perceptual-motor skills, and the identification of the most appropriate strategies to organize practice are an entire sub-discipline of sports science. For instance, there is a long history of research investigating the use of demonstrations, modeling, and feedback, among other techniques [see \citet{Farrow13} for more about expert-novice differences and developing sport expertise]. A similar approach to understanding skill in astronomy may prove useful and particularly informative for future developments.  We leave this as an opportunity for further study.

\section{Conclusion}
\label{sect:conclusion}
Current and future observatories will increasingly work within the framework of continuous survey operations.  Correspondingly, astronomy is moving from a mode of user-centered, desktop-bound, post-collection processing and analysis of data to one of continuous, interactive, real-time analysis of streaming data.  Techniques for knowledge discovery must evolve and scale to keep up with the data rate.  

For the specific case of extragalactic neutral hydrogen spectral data cube studies, we have demonstrated that the {\sc SportsCode} performance analysis software can be used as part of a process for understanding how astronomers make, or confirm, discoveries visually.  With minimal training in its use, an expert H{\i} astronomer was able to inspect a data cube encoded as a movie, and select and apply appropriate codes and annotations.  Although not intended as an astronomical analysis tool, we discerned that {\sc SportsCode} could be used alongside a more traditional solution.  In this case, {\sc SportsCode} would play the role of a visual logbook that can be used to highlight -- by drawing on the movie frames -- the locations of features of interest. Alternatively, {\em in situ}  annotation and coding of features would be a valuable addition to existing and planned visualization and analysis packages.  Through coding, graphical, text and other annotations, the reasons behind classifications as ``source'' or ``noise'' can be both preserved and subjected to further study.

With the expectation that new astronomers will have to learn visual-based discovery skills much more rapidly than previous generations, we suggest that astronomy may benefit from the adoption of a performance analysis framework.   As in the world of sports, all astronomers would benefit from specialist coaching, and elite talent must be identified and nurtured.   Performance at all aspects of analysis and visualization must undergo continuous improvement if our goal is to achieve a winning advantage.  We may not need to look at everything in the data intensive era, but we should ensure we are ready to make a discovery when we see it. 

{\sc SportsCode} licenses were provided by Swinburne University of Technology.    At various stages of this work, C.J.F. benefited from discussions with David Barnes, Sandra McClelland, and Jack Singh.  The authors thank the anonymous referee for making some very useful suggestions.


\begin{thebibliography}{}
\bibitem[Ball \& Brunner (2010)]{Ball10}
Ball, N.M.,\ Brunner, R.J., 2010, IJMPD, 19, 1049

\bibitem[Banerji et al.\ (2010)]{Banerji10}
Banerji, M., Lahav, O., Lintott, C.J., et al., 2010, MNRAS, 406, 342

\bibitem[Barnes et al.\ (2006)]{Barnes06}
Barnes, D.G., Fluke, C.J., Bourke, P.D., Parry, O.T., 2006, PASA, 23, 82

\bibitem[Bell et al.\ (2009)]{Bell09}
Bell, G., Hey, T., Szalay, A., 2009, Science, 323, No. 5919,1297

\bibitem[Berlucchi (2011)]{Berlucchi11}
Berlucchi, G., 2011, MemSAIt, 82, 225

\bibitem[Borkin et al.\ (2005)]{Borkin05}
Borkin, M.A., Ridge, N.A., Goodman, A.A., Halle, M., 2005, arXiv:astro-ph/0506604

\bibitem[Borne (2009)]{Borne09}
Borne, K., 2009, Scientific Data Mining in Astronomy. In Next Generation of Data Mining (Taylor \& Francis: CRC Press), Ch. 5, 91

\bibitem[Brown et al.\ (2004)]{Brown04}
Brown, R.L., Wolfgang, W., Cunningham, C., 2004, AdSpR, 34, 555

\bibitem[Brunner et al.\ (2002)]{Brunner02}
Brunner, R. J., Djorgovski, S. G., Prince, T. A., Szalay, A. S., 2002, Massive Datasets in Astronomy. In Handbook of Massive Datasets (Springer US), Ch. 27, 931

\bibitem[Callary et al.\ (2012)]{Callary12}
Callary, B., Werthner, P., Trudel, P., 2012, Qualitative Research in Sport, Exercise and Health, 4, 420


\bibitem[Dewdney et al.\ (2009)]{Dewdney09}
Dewdney, P.E., Hall, P.J., Schilizzi, R.T., Lazio, T.J.L.W., 2009, Proceedings of the IEEE, 97, 1482

\bibitem[Djorgovski \& Williams (2005)]{Djorgovski05}
Djorgovski, S.G., Williams, R., 2005, Virtual Observatory: From Concept to Implementation. In From Clark Lake to the Long Wavelength Array: Bill Erickson?s Radio Science, 345, 517

\bibitem[Ericsson et al.\ (1993)]{Ericsson93}
Ericsson, K. A., Krampe, R. T., Tesch-R\"{o}mer, C., 1993, Psychological review, 100, 363

\bibitem[Fabian (2009)]{Fabian09}
Fabian, A.C., 2009, arXiv:0908.2784

\bibitem[Farrow et al.\ (2013)]{Farrow13}
Farrow, D., Baker, J., MacMahon, C., eds, 2013,  Developing sport expertise: Researchers and coaches put theory into practice,  Routledge.

\bibitem[Ferrand et al.\ (2016)]{Ferrand16}
Ferrand, G., English, J., Pourang, I., 2016, arXiv:1607.08874

\bibitem[Gooch (1997)]{Gooch97}
Gooch, R.E., 1997, PASA, 14, 106

\bibitem[Hassan \& Fluke (2011)]{Hassan11}
Hassan, A.H., Fluke, C.J., 2011, PASA, 28, 150

\bibitem[Hassan et al.\ (2013)]{Hassan13}
Hassan, A.H., Fluke, C.J., Barnes, D.G., Kilborn, V.A., 2013, MNRAS, 429, 2442

\bibitem[Hatala et al.\ (1999)]{Hatala99}
Hatala, R., Norman, G.R., Brooks, L.R., 1999, Journal of General Internal Medicine, 14, 126


\bibitem[Hey et al.\ (2009)]{Hey09}
Hey, T., Tansley, S., Tolle, K. M., 2009, The Fourth Paradigm: Data-Intensive Scientific Discovery. Microsoft Research (Redmond, Washington)

\bibitem[Hollis (1938)]{Hollis38}
Hollis, H.P., 1938, Obs, 61, 301


\bibitem[Ivezi\'{c} et al.\ (2014)]{Ivezic14}
Ivezi\'{c}, \.{Z},  Connolly, A. J., VanderPlas, J. T., Gray, A., 2014,Statistics, Data Mining, and Machine Learning in Astronomy: A Practical Python Guide for the Analysis of Survey Data (Princeton, New Jersey: Princeton University Press)

\bibitem[Jonas (2009)]{Jonas09}
Jonas, J.L., 2009, Proceedings of the IEEE, 97, 1522

\bibitem[Johnston et al.\ (2008)]{Johnston08}
Johnston, S., Taylor, R., Bailes, M., et al., 2008, ExA, 22, 151

\bibitem[Kent (2013)]{Kent13}
Kent, B.R., 2013, PASP, 125, 731

\bibitem[Koribalski (2012)]{Koribalski12}
Koribalski, B.S., 2012, PASA, 29, 359

\bibitem[Lang et al.\ (2003)]{Lang03}
Lang, R.H., Boyce, P.J., Kilborn, V.A., et al.\, 2003, MNRAS, 342, 738

\bibitem[Lintott et al.\  (2008)]{Lintott08}
Lintott, C. J.,  Schawinski, K.,  Anze, S., et al.,  2008, MNRAS, 389, 1179 

\bibitem[Lintott et al.\  (2009)]{Lintott09}
Lintott, C.J.,  Schawinski, K., Slosar, A., et al., 2009, MNRAS, 399, 129

\bibitem[MacMahon \& Starkes (2008)]{MacMahon08}
MacMahon, C., Starkes, J.L., 2008, Journal of Sports Sciences, 26, 751


\bibitem[Mitchel (1858)]{Mitchel58}
Mitchel, O.M., 1858, MNRAS, 18, 261

\bibitem[Naiman (2016)]{Naiman16}
Naiman, J.P., 2016, A\&C, 15, 50

\bibitem[Norman et al.\ (2006)]{Norman06}
Norman, G., Eva, K., Brooks, L., Hamstra, S., 2006, Expertise in Medicine and Surgery. In The Cambridge Handbook of Expertise and Expert Performance (Edward Elgar Publishing), 339


\bibitem[Norris (2010)]{Norris10}
Norris, R.P., 2010, arXiv:1009.6027v1

\bibitem[Perkins et al.\ (2014)]{Perkins14}
Perkins, S., Questiaux, J., Finniss, S., Tyler, R., Blyth, S., Kuttel, M.M., 2014, NewA, 30, 1

\bibitem[Popping et al.\ (2012)]{Popping12}
Popping, A., Jurek, R., Westmeier, T., Serra, P., Fl\"{o}er, L., Meyer, M., Koribalski, B., 2012, PASA, 29, 318

\bibitem[Punzo et al.\ (2015)]{Punzo15}
Punzo, D., van der Hulst, J.M., Roerdink, J.B.T.M., Oosterloo, T.A., Ramatsoku M., Verheijen, M.A.W., 2015, A\&C, 12, 86

\bibitem[Punzo et al.\ (2016)]{Punzo16}
Punzo, D., van der Hulst, J.M., Roerdink, J.B.T.M., 2016, A\&C, 1790, 163

\bibitem[Rogowitz \& Goodman (2012)]{Rogowitz12}
Rogowitz, B.E., Goodman, A., 2012,  Proc. SPIE 8291, Human Vision and Electronic Imaging XVII, 82910W

\bibitem[Ruiz et al.\ (2014)]{Ruiz14}
Ruiz, J.E., Garrido, J., Santander-Vela, J.D., S\'{a}nchez-Exp\'{o}sito, S., Verdes-Montenegro, L., 2014, A\&C, 7, 3
 
\bibitem[Serra et al.\ (2015)]{Serra15}
Serra, P., Westmeier, T., Giese, N., Jurek, R., Fl\"{o}er, L., Popping, A., Winkel, B., van der Hulst, T., Meyer, M., Koribalski, B.S., Staveley-Smith, L., Courtois, H., 2015, MNRAS, 448, 1922

\bibitem[Szalay \& Gray (2001)]{Szalay01}
Szalay, A., Gray, J., 2001, Science, 293, 2037

\bibitem[Szalay \& Gray (2006)]{Szalay06}
Szalay, A., Gray, J., 2006, Nature, 440, 413


\bibitem[Taylor (2015)]{Taylor15}
Taylor, R., 2015, A\&C, 13, 67

\bibitem[Tyson (2002)]{Tyson02}
Tyson, J.A., 2002,  in ``Survey and Other Telescope Technologies and Discoveries'', J.A. Tyson, J.A., S. Wolff (eds),  Proc. SPIE, 4836, 10

\bibitem[van Haarlem et al.\ (2013)]{vanHaarlem13}
van Haarlem, M.P., Wise, M.W., Gunst, A.W., et al., 2013, A\&A, 556, A2

\bibitem[Vohl et al.\ (2016)]{Vohl16}
Vohl, D., Barnes, D.G., Fluke, C.J., Poudel, G., Georgiou-Karistianis, N., Hassan, A.H., Benovitski, Y., Wong, T.H., Kaluza, O., Nguyen, T.D., Bonnington, C.P., 2016, PeerJ CS, 2, e88

\bibitem[Ward \& Barker (2013)]{Ward13}
Ward, J.S., Barker, A., 2013. arXiv:1309.5821

\bibitem[Whiting (2012)]{Whiting12}
Whiting, M.T., 2012, MNRAS, 421, 3242

\bibitem[Wolfinger et al.\ (2013)]{Wolfinger13}
Wolfinger, K., Kilborn, V.A., Koribalski, B.S., Minchin, R.F., Boyce, P.J., Disney, M.J., Lang, R.H., Jordan, C.A., 2013, MNRAS, 428, 1790

\end{thebibliography}
\end{document}